\documentclass[sigplan,nonacm]{acmart}

\usepackage{tikz}
\usepackage{cleveref}
\usepackage{amsmath}

\usepackage{filecontents}

\usepackage{algorithmic}
\usepackage{graphicx}
\usepackage{textcomp}
\usepackage{xcolor}

\definecolor{indigo}{RGB}{51,0,102}
\usepackage{tcolorbox}

\usepackage{hyperref}

\usepackage[framemethod=TikZ]{mdframed}

\usepackage{xspace}
\usepackage{scalerel}
\usepackage{ctable}

\usepackage{pifont}

\usepackage{url}
\usepackage{mathtools}

\usepackage{tabularray}

\usepackage{multirow}
\usepackage{array}
\usepackage[normalem]{ulem}

\AtBeginDocument{%
  }

\copyrightyear{2026}

\usepackage{listings}

\definecolor{brightpink}{rgb}{1.0, 0.0, 0.5}

\newcommand{\fixme}[1]{\textcolor{red}{(\textbf{Fix Me}) #1}}
\newcommand{\sahm}[0]{SAHM}

\begin{document}

\title{SAHM: State-Aware Heterogeneous Multicore for Single-Thread Performance}
\author{Shayne Wadle, Karthikeyan Sankaralingam}
\affiliation{%
  \institution{University of Wisconsin - Madison}\city{Madison}\country{USA}}

\begin{abstract}
Improving single-thread performance remains a critical challenge in modern processor design, as conventional approaches such as deeper speculation, wider pipelines, and complex out-of-order execution face diminishing returns. This work introduces SAHM—State-Aware Heterogeneous Multicore—a novel architecture that targets performance gains by exploiting fine-grained, time-varying behavioral diversity in single-threaded workloads. Through empirical characterization of performance counter data, we define 16 distinct behavioral states representing different microarchitectural demands. Rather than over-provisioning a monolithic core with all optimizations, SAHM uses a set of specialized cores tailored to specific states and migrates threads at runtime based on detected behavior. This design enables composable microarchitectural enhancements without incurring prohibitive area, power, or complexity costs.

We evaluate SAHM in both single-threaded and multiprogrammed scenarios, demonstrating its ability to maintain core utilization while improving overall performance through intelligent state-driven scheduling. 
Experimental results show opportunity for 17\% speed up in realistic scenarios. These speed ups are robust against high-cost migration, decreasing by less than 1\%. Overall, state-aware core specialization is a new path forward for enhancing single-thread performance.
\end{abstract}


\maketitle

\pagestyle{plain}


\section{Introduction}
\label{sec:introduction}
Improving single-thread performance remains a critical and ongoing challenge in modern processor design. Although multicore scaling and accelerators have dominated recent advances in throughput, the performance of individual threads---often measured as Instructions Per Cycle (IPC)---remains vital for latency-sensitive workloads, legacy software, and many user-facing applications. Traditional approaches to boost IPC have focused on deepening speculation, widening pipelines, and complex out-of-order execution; however, these techniques face diminishing returns in the face of increasing complexity and energy constraints. Research ideas abound on various ways to increase IPC with better prefetch/caching for targeted program behaviors~\cite{analysis-specializedaccel, speedup-branchprediction, speedup-l1dcachemiss1, speedup-l2cachemiss, ipc-graph-data, speedup-l1icachemiss, speedup-l2cachemiss2, speedup-regfilemiss}.

Recent product trends illustrate this plateau: for example, AMD's Ryzen CPU generations have shown incremental single-thread performance improvements of roughly 5--10\% per generation, with much of the performance gain coming from improved clock speed, not fundamentally better per-cycle efficiency. These marginal gains highlight the need for new architectural strategies that go beyond conventional microarchitectural tuning. This paper proposes such a strategy to directly tackle the problem of single-thread performance.

Our approach is grounded in a set of key empirical observations. First, we find that single-threaded applications exhibit substantial behavioral heterogeneity across time. In particular, they vary widely in how they stress four key microarchitectural components: \textbf{the branch predictor, L1 data cache, L1 instruction cache, and L2 cache.} By collecting real hardware performance counter data from a state-of-the-art CPU across a wide range of applications, we demonstrate that this behavioral diversity is not only common, but pronounced. This constitutes the first contribution of the paper: an empirical characterization of fine-grained behavioral variability in modern workloads.

Second, we show that applications do not remain fixed in one behavioral mode. Instead, they transition between modes---or what we call \textbf{behavioral states}---throughout execution. We define a behavioral state as a 4-bit encoding derived from four key performance metrics, each binarized into HIGH or LOW categories based on observed distribution thresholds. This results in a total of 16 unique behavioral states. Different applications not only visit different subsets of these states, but also dwell in them for variable intervals and transition between them at different rates. These findings suggest that single-threaded performance could be enhanced by adapting the underlying hardware to the current behavioral state.

This observation leads to our third and central contribution: the design of a new type of processor architecture that we call \textbf{SAHM}---\textit{State-Aware Heterogeneous Multicore}. In SAHM, we construct a multicore processor where each core is specialized for one or more of the 16 behavioral states. Rather than attempting to build a one-size-fits-all core that includes all possible optimizations (e.g., a deep branch predictor, aggressive prefetchers, large instruction and data caches), we partition these features across cores based on their relevance to specific states. At runtime, we monitor the application's state and migrate it to the core best suited for the current behavior. This strategy enables composability of microarchitectural enhancements without incurring the area, power, and \textit{complexity penalties of incorporating all features into a single core}. The key design questions then become: how to detect the current state, when to trigger migration, and how to manage the cost of migration.

We further extend this idea to a realistic multi-programmed scenario, which addresses concerns about core underutilization. In a modern cloud environment, it is common to co-locate multiple independent workloads on the same processor. In such settings, SAHM's specialized cores are always busy, as different workloads occupy different behavioral states at different times. We show how an intelligent runtime scheduler can orchestrate the migration of threads across cores to maximize system-wide performance, without leaving specialized cores idle.

We also recognize practical considerations such as virtualization, cloud isolation, and security. SAHM is compatible with modern virtualization frameworks and can be integrated into hypervisors or OS-level schedulers. Migration decisions can be made within tenant boundaries to maintain isolation guarantees. Furthermore, performance counters used for state detection are already widely virtualized in commodity processors, and migration policies can be designed to respect container and VM boundaries.

\textbf{Relationship to Prior Work.} SAHM differs fundamentally from prior heterogeneous processor designs such as ARM's big.LITTLE and related single-ISA heterogeneous multicore architectures. big.LITTLE focuses on trading performance for energy efficiency: LITTLE cores are deliberately simplified to save power, while big cores are designed for high performance. In contrast, SAHM is designed to \emph{improve} performance across the board by targeting microarchitectural behaviors rather than coarse-grained application phases. Table~\ref{tab:compare} summarizes the key differences.

\begin{table*}[tbp]
\centering
\begin{tabular}{@{}lll@{}}
\toprule
\textbf{Aspect} & \textbf{big.LITTLE} & \textbf{SAHM} \\ \midrule
Core Design Goal & Energy-efficiency vs. Performance & Performance specialization by behavior \\
Granularity & Coarse (application phase) & Fine (100ms epoch) \\
State Awareness & None or static hints & Performance counter-driven \\
Migration Trigger & OS heuristics & Runtime behavioral state detection \\
Core Utilization & Some idle in steady state & Fully utilized in multiprogrammed workloads \\
\bottomrule
\end{tabular}
\caption{Comparison of SAHM with big.LITTLE}
\label{tab:compare}
\end{table*}

\textbf{In summary, this paper makes the following contributions:}
\begin{itemize}
    \item We introduce a new behavioral state taxonomy derived from performance counter measurements and show that applications exhibit diverse and time-varying microarchitectural demands. Section~\ref{sec:motivation} and \ref{sec:characterization}
    \item We propose SAHM, a heterogeneous multicore architecture in which each core is specialized for one or more behavioral states, and threads migrate at runtime to the best-fit core. Section~\ref{sec:design}
    \item We evaluate SAHM both for single-threaded applications and for multi-programmed systems, showing substantial performance benefits under realistic assumptions. Section~\ref{sec:results}
    \item We discuss implementation considerations, including migration policies, prediction strategies, and system-level integration.
\end{itemize}

\section{Overview and Motivation of Heterogeneity of Applications}
\label{sec:motivation}
In this section we present motivation for state awareness and an overview of a state aware heterogeneous multicore system.

\subsection{Understanding and Quantifying the Opportunity}
\label{sec:motivation_opportunity}
To understand the opportunity for improving single-thread performance through microarchitectural specialization, we begin by studying how modern applications behave at runtime. Our investigation is motivated by the hypothesis that even standard benchmark applications---such as those from the SPEC suite---\textit{impute different demands on the processor during different parts of execution}. If these demands can be captured and used to inform hardware behavior, we can potentially design architectures that are dynamically tailored to the application's needs.

We use performance counters available on modern CPUs to measure four key microarchitectural metrics:
\begin{table}[H]
    \centering
    \begin{tabular}{c|c}
         Metric & PMC Values \\ \hline
         \multirow{2}{*}{Branch Misprediction Ratio} & Branch Misprediction Count / \\
         & Branch Instruction Count \\ \hline
         \multirow{2}{*}{L1I Cache Miss Rate} & L1I Cache Miss Count /\\
         & Instruction Count\footnote{This is a non-programmable counter} \\ \hline
         \multirow{2}{*}{L1D Cache Miss Ratio} & L1D Cache Miss Count /\\
         & L1D Cache Access Count \\ \hline
        \multirow{2}{*}{L2 Cache Miss Ratio} & L2 Cache Miss Count /\\
        & L2 Cache Access Count \\
    \end{tabular}
    \caption{Performance monitoring metrics and the values used to calculate them. $^1$This is a non-programmable counter.}
    \label{tab:metrics_values}
\end{table}
Using modern tools (e.g., Linux \texttt{perf}, \texttt{likwid}\cite{tool-likwid}), we collect all four metrics simultaneously with negligible overhead (\textasciitilde0.5\% application slowdown), enabling fine-grained tracking without affecting program behavior. 

To structure this data, we classify each metric as either HIGH or LOW, based on cutoffs derived intuitively or from the empirical distribution (e.g., below the 50th percentile is LOW, above the 50th is HIGH). This results in a 4-bit encoding of the application's current behavior, defining one of 16 possible \textbf{behavioral states}. This state space is determined by static cut-offs unlike related work that uses performance metric stability during execution to define application phases. We sample the application at regular 100ms intervals---\textit{epochs}---and label each epoch with its corresponding behavioral state. Thus, an application’s execution trace can be viewed as a time series over this 16-state space.

We analyze the behavioral state traces (details in Section 3) of a variety of single-threaded applications and make several key observations:

\textbf{(a) State Coverage Across Applications.} Different applications occupy different subsets of the state space. For each application, we plot the percentage of time spent in each of the 16 states, showing that state distribution is highly non-uniform and workload-specific. Some applications are dominated by branch-heavy phases, while others exhibit memory intensity or front-end bottlenecks.

\textbf{(b) State Transition Dynamics.} Applications do not remain in a single state but transition across states at different frequencies. Using per-application heatmaps of state-to-state transitions, we observe a variety of transition behaviors---some applications flip rapidly between states, while others maintain longer intervals within a dominant state.

\textbf{(c) Interval Length Variability.} We define a state interval as a consecutive sequence of epochs during which the application remains in the same behavioral state. We compute the average interval length per state for each application, showing that some states are stable over hundreds of milliseconds, making them strong candidates for state-specific optimization.

\textbf{(d) Opportunity for Specialization.} For each state, we analyze how much time each application spends in it. This cross-application view reveals three insights:
\begin{itemize}
    \item No single state dominates execution across all applications. This reinforces the idea that different workloads stress different microarchitectural components.
    \item State 0---the ideal state where all four metrics are LOW---accounts for only 8\% of runtime on average. This highlights the potential for performance gain: even modest speedups in the remaining states can lead to meaningful improvements.
    \item For instance, if 30\% speedup is possible in the non-state-0 regions, overall application speedup could approach 22\%---a significant figure in the context of single-thread IPC improvements.
\end{itemize}

Each of these findings is expanded to its own subsection at the end of this section.
These findings strongly motivate the idea that instead of designing one general-purpose core to handle all scenarios, we should build multiple specialized cores, each tuned to perform well for a subset of the behavioral states. SAHM implements this vision through a practical and scalable hardware-software co-design approach.

\subsection{\sahm{} overview}
There are two central ideas to systems using State Aware Heterogeneous Multicore: 1) A CMP with diverse specialized cores and 2) A performance monitoring mechanism within the scheduler. 

\begin{figure}
  \centering
  \includegraphics[width=\linewidth]{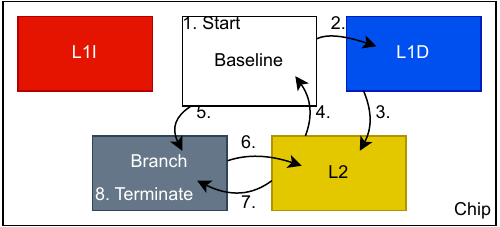}
  \caption{The canonical configuration of a \sahm{}  system. Each core except the baseline is specialized for the listed component. An example program migration pattern is included.}
  \label{fig:canonical_sahm}
\end{figure}  

Point 1 is that a each core allocates more area and power to one component - different components for each core. Figure~\ref{fig:canonical_sahm} depicts a canonical configuration in which four of the five cores have been specialized to a different component each. Due to the higher allocation, more complex designs are possible providing speed up versus the baseline component. Yet, this speed up is attainable only if the program is placed on the core matching the stressed component. 

Thus, point 2 is a scheduler that uses performance monitoring to influence which program-core assignment. The scheduler first gathers the characteristic metrics from the previous OS timestep. Next, it decides based on the metrics if the program should migrate. If so, then migration occurs to a suitable core if load-balancing conditions are met. Finally, the scheduler chooses a program from the core-local queue and launches it. Figure~\ref{fig:canonical_sahm} also shows an example program migrating between cores over time.

\subsection{Alleviating Concerns}

In proposing SAHM, it is important to address several potential concerns or misconceptions that readers may have:

\textbf{Isn’t this just adding complexity for no gain?} Not at all. The goal is to reduce system-wide complexity by avoiding the integration of all expensive features (e.g., large branch predictors, deep prefetchers, massive caches) into every core. By distributing these features across specialized cores and migrating applications to the best-fit core based on current behavior, we achieve better performance-per-area and performance-per-watt.

\textbf{Why not just make all cores smarter?} Adding every optimization into every core is cost-prohibitive in terms of area, power, and verification effort. Worse, many optimizations conflict: an aggressive data prefetcher may hurt icache locality, or a large predictor may increase access latency. SAHM allows for decoupling and composability.

\textbf{Aren’t migrations expensive?} Migration is rare and coarse-grained (every 100ms epoch), and we assume a modest 5ms cost. Our results show that state intervals are often long enough to amortize this cost, and even reactive migration policies are effective.

\textbf{How is this different from big.LITTLE?} big.LITTLE trades performance for energy efficiency using static cores. SAHM uses all high-performance cores, each tuned for different runtime behavior. Migration is driven by microarchitectural signals, not OS policies. The goal is performance, not energy reduction.

\textbf{What about OS, VMs, or security concerns?} SAHM is compatible with virtualization. Modern OSes already support thread migration. Migration policies can be constrained within VM or container boundaries. Performance counters are virtualizable and can be used without breaking isolation.

By proactively addressing these issues, we aim to clarify the feasibility and relevance of the SAHM architecture for real-world deployment.

\section{Characterization}
\label{sec:characterization}
The section delves into the program characterization including data gathering method, analysis of the collected data, and discussion of the analysis. 

\subsection{Methodological Details}
Our measurements are collected on a Golden Cove microarchitecture, representative of recent Intel client-class CPUs. This platform supports concurrent tracking of up to seven programmable performance values using hardware performance counters. To overcome the hardware limitation on simultaneously monitoring more than seven values, modern tools like Linux \texttt{perf} use a technique called \textit{multiplexing}, wherein sets of counters are time-sliced during execution. However, in our study, we restrict ourselves to seven key values, see table~\ref{tab:metrics_values}, which can all be captured simultaneously without needing multiplexing, thereby ensuring high fidelity of measurement~\cite{counters-fuse,counters-crossplatform,counter-trusted1,counter-trusted2}. We measure all of the SPEC CPU 2017 benchmark suite.

We choose a sampling interval of 100ms after empirical calibration. While finer-grained intervals (e.g., 10ms) are possible, we observed that they introduce measurement instability and noise due to context switches, sampling overheads, and limitations of event delivery latency. At 100ms, the counter values are both stable and precise, yielding nearly 100\% accuracy as validated through cross-referencing with aggregate counter values and controlled experiments. Some benchmarks execute in less time than 100ms and are excluded from further study. This interval is also coarse enough to amortize migration costs yet fine enough to capture meaningful behavioral variation. 

While our experiments are limited to Golden Cove due to platform availability and infrastructure readiness, our methodology is entirely portable. ARM and AMD processors offer comparable performance monitoring capabilities, and the SAHM approach---classifying behavior and specializing hardware accordingly---can be directly adapted to those platforms. We view this study as a proof-of-concept for a broader architectural direction that is applicable across vendors.

\subsection{State Coverage Across Applications}
\paragraph{Goal} Show that applications occupy different subsets of the behavioral state space statically determined by cut-offs. Higher variability between applications decreases the likelihood that they stress the same component at the same time when in a workload together. As the process of defining the state space includes binning, which states applications occupy is dependent on the cut-offs that determine the binning. 

\paragraph{Experiment} First, analyze the gathered traces to determine a few empirical cut-offs and include one set of cut-offs from intuition based on prior architecture experience. These are listed in Table~\ref{tab:cutoffs}. The empirical cut-offs are percentiles from the gathered traces - i.e. 50\% is the medians, 50th percentile, of each metric.
\begin{table}
    \centering
    \begin{tabular}{c|c|c|c}
         Metric & 25\% & 50\% & Intuitive\\ \hline
         branch mispred. & 0.03\% & 0.34\% & 1\% \\
         L1I miss (MPKI) & 0.004 & 0.009 & 1 \\
         L1D miss & 0.5\% & 0.99\% & 2\% \\
         L2 miss & 3.64\% & 18.47\% & 10\%
    \end{tabular}
    \caption{Three candidate cut-offs for metric binning.}
    \label{tab:cutoffs}
\end{table}

\begin{figure}
  \centering
  \includegraphics[width=\linewidth]{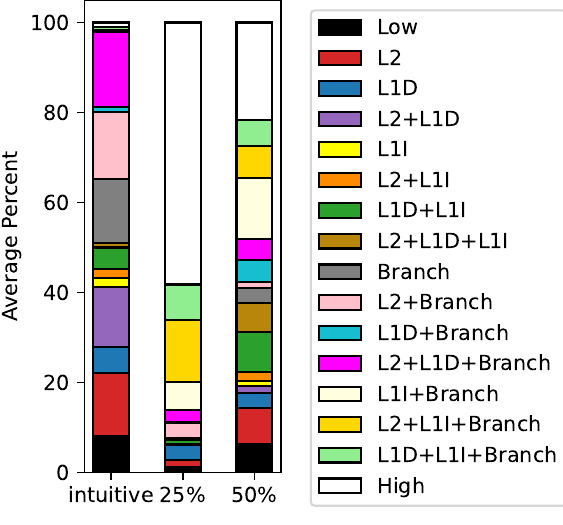}
  \caption{The portion of an application that the application is in a state averaged across the SPEC 2017 benchmarks. The intiutive cut-offs captures the most diversity.}
  \label{fig:cutoff_state_distribution}
\end{figure}

Then, calculate the portion of time spent in each state on average for each set of cut-offs. Figure~\ref{fig:cutoff_state_distribution} shows this breakdown. Each state is labeled by the combination of components stressed in that state; for example, the ``L2+Branch'' state has both the L2 miss ratio and the branch mispredict ratio as HIGH which identifies these components as stressed. One expectation of this state space is mostly mutually exclusive behavior - L2 HIGH makes sense to overlap either L1 cache metrics being HIGH. This is the case for the intuitive cut-offs; however, both of the others have a large portion of highly combined component stresses. The 25\% cut-offs have more than half of an average application spent stressing the entire core -- this does not reflect reality. Another expectation is some amount of the Low state due to the cache starting cold and paging in larger data sets. This is confirmed as 8\% of an average application is spent in the Low state for the intuitive cut-offs.

\begin{figure}
  \centering
  \includegraphics[width=0.9\linewidth]{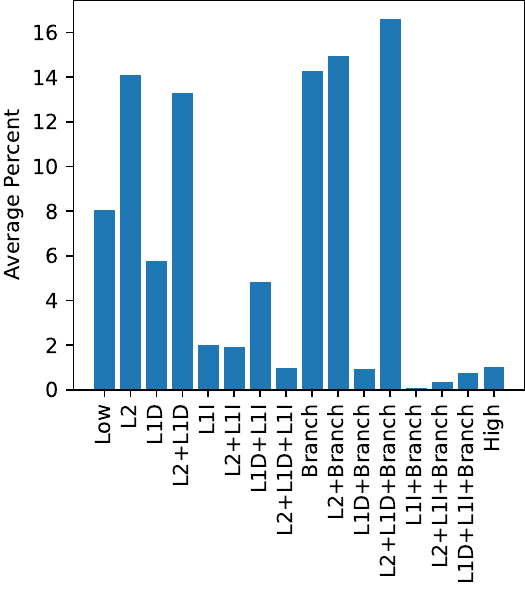}
  \caption{The portion of an average application spent in a state with intuitive cut offs. The majority of states are visited.}
  \label{fig:intuitive_cutoff_state_distribution}
\end{figure}

Figure~\ref{fig:intuitive_cutoff_state_distribution} provides another look at the intuitive column by unstacking the bars. The most occupied states are much easier to see: Low, L2, L2+L1D, Branch, L2+Branch, and L2+L1D+Branch. The overlap between data memory and branch prediction could be that branch prediction is causing cache pollution or that the data brought in from cache misses does not align with the current pattern learned by the branch predictor.

\begin{figure*}
  \centering
  \includegraphics[width=0.9\textwidth]{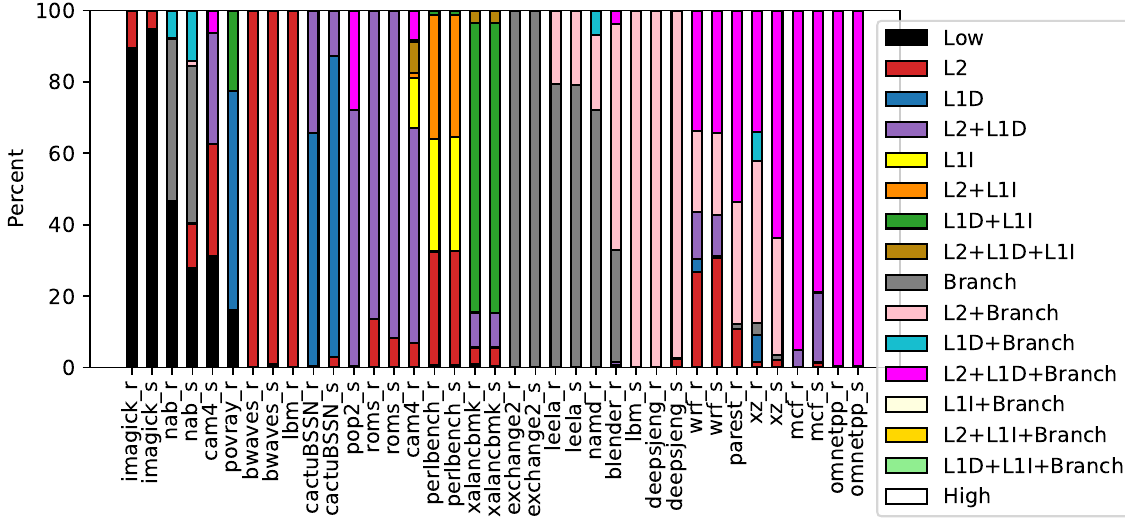}
  \caption{The percent of runtime spent in each state by application. The breadth of behaviors stands out. }
  \label{fig:intuitive_by_benchmark}
\end{figure*}

Figure~\ref{fig:intuitive_by_benchmark} breaks the intuitive column into a stack for each benchmark. It is apparent that the programs show a wide range of state occupations. Physics simulations like pop2, roms, and cactuBSSN all utilize a large working set that does not fit into the first level of cache. This is reflected in our results, as each stresses the data memory via L1D cache and L2 cache. Imagick spends the majority of its time in the LOW state - making it a poor candidate for SAHM.

\paragraph{Key takeaways} \textit{i) Different applications occupy different states in the statically determined state space.
ii) The intuitive cut-offs show the most diverse set of state occupation and more closely match how a core functions. 
iii) Almost all of the execution time is spent in states where one of the 4 key metrics is HIGH.
iv) For the rest of this work, the intuitive cut-offs are used to define the state space.}

\subsection{State Transitions}
\paragraph{Goal:} We want to understand how long applications stay in a state, when they transition, are transitions highly correlated with the originating state (i.e. out-degree of a state in this graph of transitions), and how application-dependent this behavior is. The more dynamic and application-dependent the more difficult for compilers, software transformations, or static hardware policies to exploit this.

\begin{figure}
  \centering
  \includegraphics[width=\linewidth]{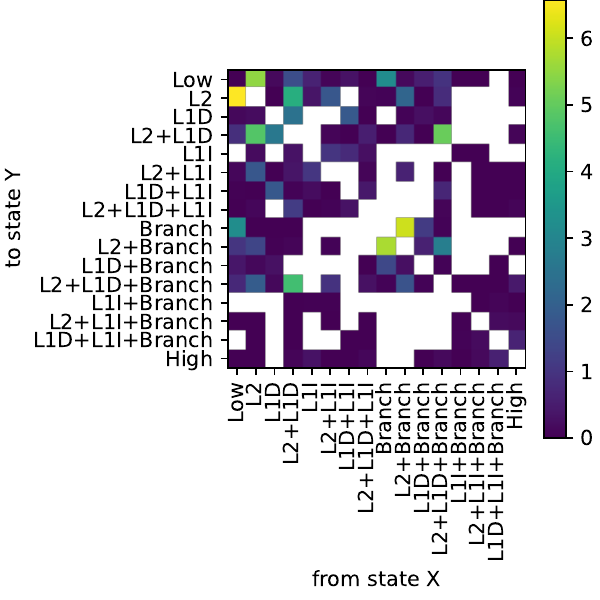}
  \caption{The portion of the total number of transitions on average. The diagonal has been removed and white cells indicate transitions that were not seen in our study. There is no overwhelming outliers that should be designed for.}
  \label{fig:intuitive_trans_dist}
\end{figure}

\paragraph{Experiment:} Analyze gathered traces for state transitions - for each epoch in a trace, examine the previous epoch and record the edge between states. Count the total for each such edge. Figure~\ref{fig:intuitive_trans_dist} depicts the heatmap of the average percent of the total number of transitions for each transition. If a transition never occurred during our analysis, then the cell for the transition is white. The diagonal, the state transitions that stay in the same state, has been removed from the figure. Transitions that stay in the same state dominate with 84\% of epochs staying in the same state as the previous epoch.

\paragraph{Analysis:} First, this confirms the expectation that transitions do occur. Programs usually go through phases and prior works have confirmed this. Next, more used transitions are between states that have more time spent in them. This is beneficial as it means that capturing the states most resided in also captures the most frequently used transitions. Furthermore, the figure is not completely symmetric across the diagonal. This indicates that order that states occur is non-trivial: states do not merely return to Low. More rigorously: the set of states and their likelihood to transition to a given state is different than the set of states and likelihood of transitions from that given state. One example of this is Low to L1I HIGH does not occur yet L1I HIGH to Low does occur.

\paragraph{Key findings:} \textit{State Transitions occur and are not uniform between applications (i.e. how an application enters state L2 HIGH is not always from L1D HIGH)}

\subsection{Interval Length Variability}
\paragraph{Goal:} We have seen transitions do occur; what is the length of the intervals between transitions that change state? We define a state interval as a consecutive sequence of epochs during which the application remains in the same behavioral state. Shorter intervals are more difficult to exploit for performance gain - the transient nature means that any adjustment to the state could take so long the state has passed by the time the adjustment is in place.

\begin{figure}
  \centering
  \includegraphics[width=0.9\linewidth]{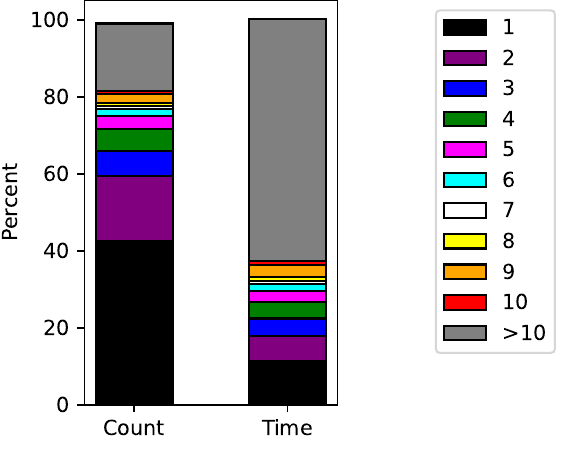}
  \caption{Analysis of intervals - strings of the same behavioral state. In count, 1 epoch intervals dominate; on the other hand, long intervals provide ample time to amortize migration overhead.}
  \label{fig:interval}
\end{figure}

\paragraph{Experiment:} Using the gathered traces, we calculate the interval length through examining the current epoch to the previous epoch like in the transition experiment. Figure~\ref{fig:interval} presents this calculation breaking down both the total count and the total time for an average application. The extremes cover the majority of the data: The short intervals make up most of the overall number of intervals. On the other hand, long intervals make up the majority of the time spent. This alerts us to a pitfall for exploiting these intervals - a program may ping-pong between two specialized cores if the program is only short intervals. Yet, there is plenty of time to amortize overhead when a long interval is reached. These factors indicate that some sort of inertia, keeping migrated programs on the assigned core for a period of time, should be included in a system that adjusts for state behavior. 

\paragraph{Key Findings:} \textit{The average interval duration provides ample opportunity to migrate based on the dynamic behavioral state, therefore extracting efficiency and/or speedup.}

\subsection{Summary}
This characterization has shown that even though the state space is determined by the cut-offs, applications occupy a swath of the space. This occupation is highly dependent on the application. Transitions between states are uniformly used and usually occur either immediately or distantly. The long intervals have applications reside in a handful of states for more than half their runtime. It is clear that specializing for these states will improve performance.

\section{SAHM Design}
\label{sec:design}
This section details the designs required for SAHM. These include the architecture of a heterogeneous multi-core, the created configuration space, and the scheduling policy. 

\subsection{Single Core Architecture}
Current designs focus on two points: designed for performance, big or P cores, or designed for energy efficiency, LITTLE or E cores. Focusing on just performance, the monolith cores are generalists - do everything mediocre. In this era of multicore, this is wasteful. Individual cores can specialize, contributing their slice of the behavioral state space to the CMP as a whole. In a \sahm{} system, each core attributes most of the area and power budget to one component. The next Section~\ref{sec:related_components} provides examples of components that lead to speed up when there is budget allocated for them. 

\subsection{\sahm{} Architecture}
\label{sec:architecture}
Individual specialized cores by themselves cannot compete with the monolith cores. In this case, the system is greater than the sum of its parts: the specialized cores cover the deficiencies of each other. Figure~\ref{fig:canonical_sahm} depicts one conservative general purpose CMP. It includes a general core and specialty cores for each component. As a program executes it will be placed onto the core that best suits its current operating needs. A \sahm{} system does not necessarily have a specialized core for each component; some may be left out so others may be duplicated. 

This introduces a new architecture space: the number and type of specialized cores on a CMP. A chip designer can use analysis akin to Figure~\ref{fig:intuitive_by_benchmark} and knowledge of the workload the chip will run, if known, to pick the most apt configuration within the space. For example, cactuBSSN, pop2, and roms are all physics modeling programs and all stress the data memory. A chip designer would place more cores specialized to handle large working sets into  CMP that will execute these programs. This work examines hypothetical configurations as a primary exploration: those with 1 baseline core and a number of uniquely specialized cores:
\begin{itemize}
    \item 1 specialty core results in 4 configurations
    \item 2 specialty cores results in 6 configurations
    \item 3 specialty cores results in 4 configurations
    \item and 4 specialty cores results in 1 configuration
\end{itemize}
Furthermore, we assume potential speedups gained by each specialty core of 10\%, 20\%, and 30\%. The total number of configurations is 256 including a baseline chip with no specializations. 
These assumed speedups are seen in many recent microarchitecture works. We provide an example for each key metric that achieves between 10\% and 30\% speed up.

\label{sec:related_components}

\paragraph{Branch Misprediction:} Eyerman et al. achieved an average 29\% increase in performance through selectively flushing instructions after mispredicted branches \cite{speedup-branchprediction}.

\paragraph{L1 instruction cache miss:} The entangling prefetcher achieves 10\% speedup in less size than the compared instruction prefetchers \cite{speedup-l1icachemiss}. Also, increasing the size of the L1 instruction cache is viable.

\paragraph{L1 data cache miss:} Berti: an Accurate Local-Delta Data Prefetcher \cite{speedup-l1dcachemiss1} and Register file prefetching \cite{speedup-regfilemiss} both provide 5\% improvements. Berti accomplishes this by selecting the best local-deltas. Register file prefetching orchestrates the data pipeline in an out-of-order system to prefetch nearly half of load requests to the register file.

\paragraph{L2 cache miss:} Wu et al. achieve approximately 30\% speed up by the proposed Triage prefetcher that filters for important meta-data and sometimes uses last level cache to store additional meta-data \cite{speedup-l2cachemiss2}. Saglam et al. achieve 30\% to 130\% speed up in HBM2 memory systems by using a proposed aggressive prefetcher \cite{speedup-l2cachemiss}. In DDR systems, the prefetcher switches to a conservative mode with no performance gain.

Re-implementing these published microarchitecture techniques and running them at cycle-level simulation is infeasible and unnecessary for 100ms epochs. Instead, SAHM builds on this established work, composing them in a novel way making it bigger than the sum of its parts. Essentially, our system can be thought of as implementing each of those specific ideas into their own core with only that specialization. 

Furthermore, these works were invented and evaluated through the lens of generalist cores: all applications execute their whole duration on the proposed components. Our deployment of them in targeted phases - the most suitable portions of applications execute on each component - should mean our speedups are easily viable.

\subsection{Software: Scheduling}
A chip using only specialized cores theoretically gains performance. Unfortunately, chips do not exist in a vacuum: current schedulers, like CFS~\cite{CFS-CFS}, do not acknowledge asymmetry nor specialization~\cite{CFS-ACFS,CFS-Colab}. All parts of the system must work together to improve performance. Thus, we have designed a simple greedy scheduler; the algorithm is seen in Listing~\ref{alg:local_sched}. Three pieces enable the scheduler to use specializations. The first is a multi-queue structure that includes a mapping of specializations to cores. The second is a performance analysis section to determine behavioral state and suitable cores. The third is an inertia component to ensure that the numerous short intervals do not cause continuous migrations of a program. The inertia prevents migration for a certain number of schedulings after migration to a new core. An oracle version of this scheduler has knowledge of which states the program will enter in the next execution timespans. 

\begin{lstlisting}[caption={SAHM scheduler incorperating state analysis and program-core inertia}, label={alg:local_sched}]
if current_program.inertia > 0:
  # do nothing as the program has 
  # migrated too recently
  decrement current_program.inertia
  launch_from_local_queue()

perf_counts = get_PMU_counts()
state = calculate_state(perf_counts)
if not specialty_matches(
    current_core.specialty, state):
  # choose a new core
  core = is_core_available_for(state)
  program_handle = current_core. \
                 dequeue(current_program)
  if core is null:
    # load balance if not specialty
    core = most_idle_core()
  core.queue(program_handle)
  # queueing also sets program inertia
launch_from_local_queue()
\end{lstlisting}

\section{Evaluation Framework}
\label{sec:method}
This evaluation is based on a first-order model and a simulator. For the model, a chip has \textit{n} specialized cores, each defined by the specialized component and the provided speed up. The workload is selected from the SPEC 2017 CPU benchmark suite. A program is represented by a trace of its gathered performance values. Once a benchmark concludes it restarts; this keeps the mix of applications and behaviors diverse over time. A test executes for a fixed time: the length of the longest benchmark in the executing workload. 

This model is developed into a trace driven simulator in order to test the configurations, the scheduler, and the impact of core contention and migration cost. The simulator first initializes the chip and the scheduler data structures. Each benchmark starts at the beginning of its execution trace and has its progress tracker set to zero. Then, the simulator enters the main loop: first, invoke the scheduler on all cores individually making migrations as necessary. Then, perform a timestep of 10ms for each program currently assigned to a core - not waiting in the queue for a core. If a program is assigned to a core whose specialization matches the current state, then the program advances further than on a baseline core; it is sped up. Due to the representation of the workload, non-uniform scheduling events like system calls are not included.  

\subsection{Metrics}
During evaluation, we compute:
\begin{itemize}
    \item \textbf{System-level Speedup}: total throughput relative to a baseline with an equal number general-purpose cores.
    \item \textbf{Per-Application Speedup}: how each app benefits or suffers under the scheduling policy.
    \item \textbf{Migration Rate}: average number of swaps per second across the system.
    \item \textbf{Epoch Utilization}: fraction of epochs where apps run on matching specialized cores.
\end{itemize}

\section{Results}
\label{sec:results}
Our results can be viewed in two lights: per benchmark and per configuration. 
First, a limited study into the per benchmark speed up is presented. 
Second, we delve into the breadth of potential speed ups per benchmark. 
Third, a general chip and a few specialized chips are examined. 
Fourth, we factor in realistic constraints like migration cost into simulation. The results demonstrate the \sahm{}  system provides speed up even in extreme cases.
Finally, included in an appendix is the whole design space of configurations and their speed ups. There are too many data points in the space to lead the reader through; we select the most interesting for this section. 

\subsection{Single Application Limit Study}
\paragraph{Goal:} What is the limit of the SAHM system in an idealized scenario?

\begin{figure}
  \centering
  \includegraphics[width=0.5\textwidth]{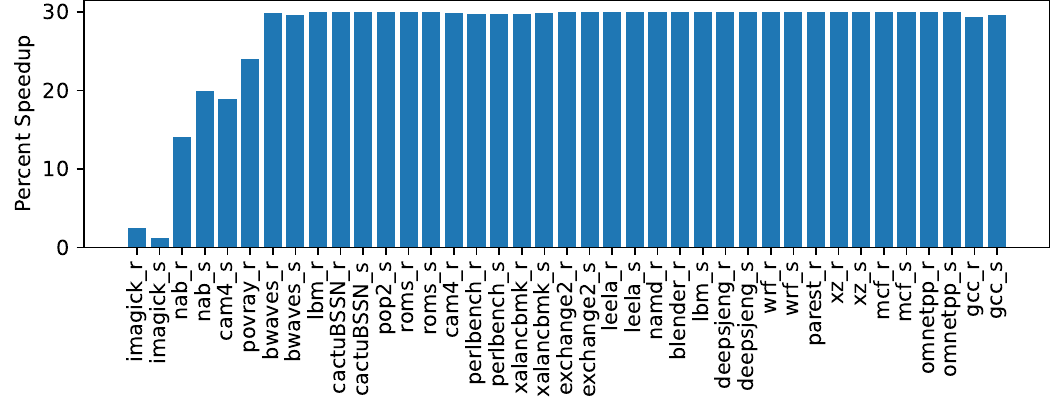}
  \caption{The maximum possible idealized speed up by benchmark}
  \label{fig:bmk_ideal_real}
\end{figure}

\paragraph{Experiment:} We model 5 cores in the canonical configuration: 1 baseline core and 4 specialized cores. The specialized cores provide 30\% speed up when an application is in a behavioral state with the specialized component being stressed. Each benchmark is individually run on the system with no migration cost. This means there is no contention for cores. The speed up achieved by each benchmark is shown in Figure~\ref{fig:bmk_ideal_real}.  

\paragraph{Key Takeaway:} \textit{The majority of benchmarks achieve high speed up and return on investment.}

\paragraph{Analysis:} These results mirror the portion of time outside of the Low state in the characterization results. Imagick has poor speedup as the vast majority of its time is spent in the Low state. Similarly for nab and cam4-s. Overall, this confirms that there is opportunity to increase performance.

\subsection{Benchmark Breadth}
\paragraph{Goal:} The size of the design space is difficult to grapple. First, we use the lens of application speedup to understand the distribution provided by other chip configurations that are composed using different number of cores, and different speedups provided by the cores. 

\begin{figure*}
  \centering
  \includegraphics[width=0.9\textwidth]{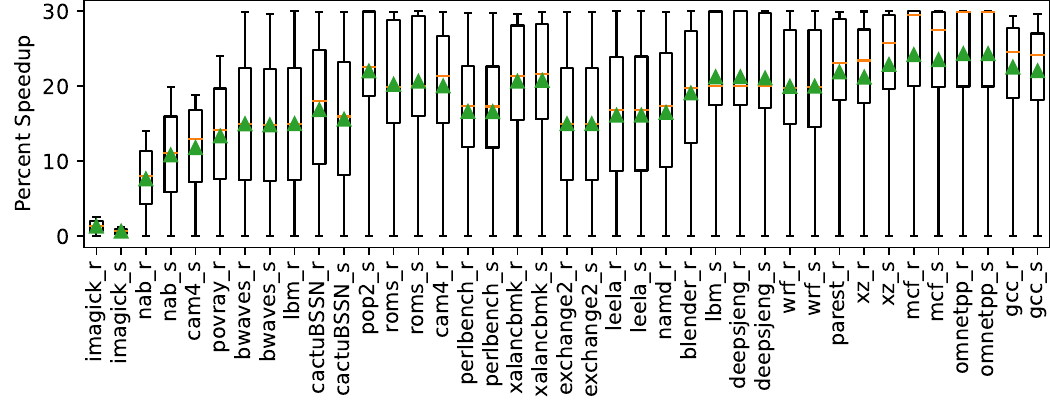}
  \caption{Distribution of percent speed up by application; the configurations create the distribution. The green triangle is the average while the orange line is the median. The box is the 25\% and 75\% percentile. The whiskers cover the entire range.}
  \label{fig:bmk_speedup}
\end{figure*}

\paragraph{Experiment:} Each benchmark is simulated on a range of configurations. There is no contention or migration cost so that we can understand the potential of the design space. The configurations are as described in Section~\ref{sec:architecture} that results in 255 configurations - we do not include a baseline configuration here in order to tell if an application does not achieve performance from a configuration that includes specialization. The distribution per benchmark is displayed in Figure~\ref{fig:bmk_speedup}. The whiskers are the entire range which explains how the top whisker extends to the ideal speed up as seen in Figure~\ref{fig:bmk_ideal_real}. The green triangles show the mean and the orange line is the median. These along with the box of the 25th and 75th percentile show the skew of the distribution. 

\paragraph{Key Finding:} \textit{The majority of configurations provide speed up to the majority of applications.}

\paragraph{Analysis:} The average speed up is 15\% across applications. Additionally, the distributions are skewed toward higher speed up seen through the longer distance to the top of the box compared to the bottom of the box from the median line. This indicates that the majority of configurations outperform the average speed up. Furthermore, the average 25th percentile is over 10\% speed up. This demonstrates that even in poorly aligned configurations - configurations that are missing specialization for the behavioral state the program resides in most - applications still benefit from specialization.

\subsection{Generalist vs Specialized}
\label{sec:general_special}
\paragraph{Goal:} Considering the vast design space, does incremental progress on a single core outperform a system of specialized cores in single-threaded performance? 

\begin{figure}
  \centering
  \includegraphics[width=0.9\linewidth]{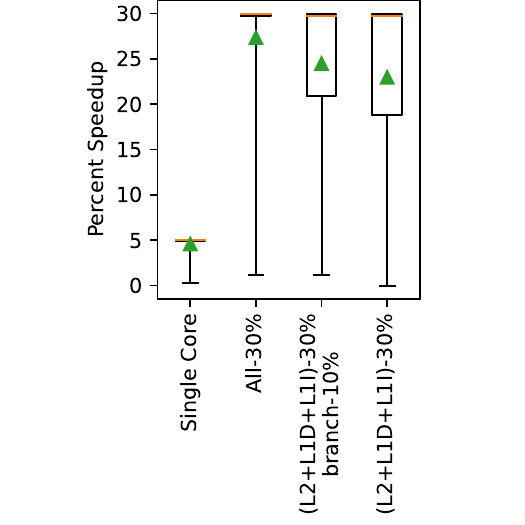}
  \caption{Distribution of speed up by configuration. Applications create distribution. Green triangle is mean while orange line is median. Box is the 25\% and 75\% percentile. Whiskers are the entire range.}
  \label{fig:config_speedup}
\end{figure}

\paragraph{Experiment:} We simulate a representative sample of configurations as well as a single core that has 5\% improvement overall. Each benchmark is individually simulated on a configuration with no migration cost. This provides the best opportunity for single-threaded speed up to every configuration. The representative configurations are canonical 5 core systems with 30\% speed up for each specialized core except for the core specializing the branch predictor. In Figure~\ref{fig:config_speedup} we provide the configuration with 30\% speed up branch core for context under 'All-30\%' while also showing systems with the branch core providing 10\% and 0\% speed up. 

\paragraph{Key Finding:} \textit{Specialized systems for the majority of applications outperform the best an incremental approach can provide.}

\paragraph{Analysis:} The 25th percentile for each of the specialized systems is, at minimum, 3x the best speed up the single core achieves. Even for the bottom quarter of applications where specialization is not fully compatible, the speed up enabled by specialization outperforms the single core.

\subsection{Realistic Simulation}
\paragraph{Goal:} Now that we understand the theoretical speed ups of the design space, how effective is the system under real world constraints like migration cost and contention?

\begin{figure*}
  \centering
  \includegraphics[width=0.85\textwidth]{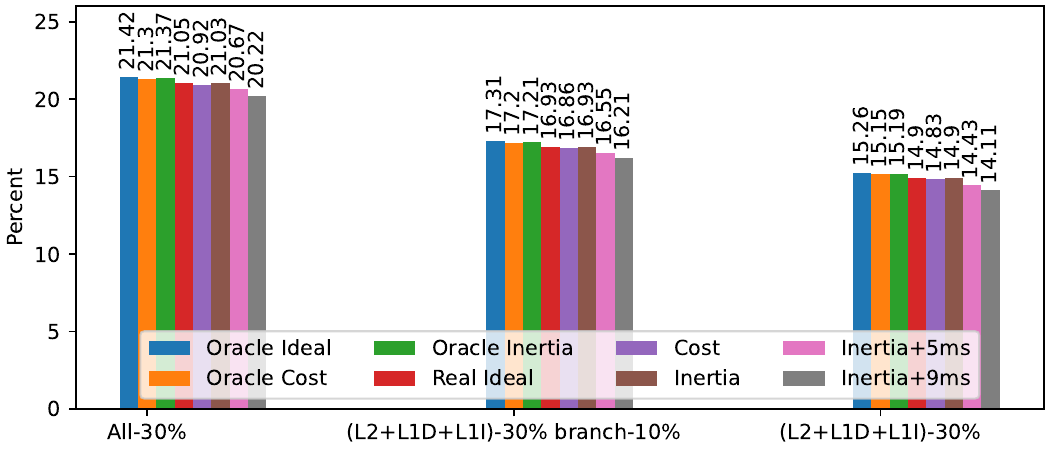}
  \caption{Average application speed up achieved by each configuration under various constraints. `Oracle' uses an oracle scheduler. `Ideal' has no migration cost. `Cost' is 1ms per migration. `Inertia' is 1ms per migration with the inertia scheduler. Inertia+time uses the inertia scheduler while increasing the migration cost to the list time.}
  \label{fig:simulate_real}
\end{figure*}

\paragraph{Experiment:} We simulate a scaled up version of the canonical configuration: it has 7 baseline cores and 8 of each specialized core. 39 cores is chosen as there are 39 benchmarks to be run on the system. This drastically simplifies results to one workload that captures the entire diversity of application behavior instead of needing multiple plots explaining and showing the results for a dozen workload mixes. These results scale to a 32 or 64 core system with equal numbers of specialized cores because the workload mixes for these systems will comprise the same behaviors. Therefore, the effects of core contention will not change.
We analyze the same specialized configurations presented in the previous Section~\ref{sec:general_special}. From an ideal system using an oracle scheduler, we introduce migration cost, realistic scheduling, and the inertia component to the scheduler. Figure~\ref{fig:simulate_real} shows the speed ups for each of 8 constraint configurations. `Oracle' indicates employ of a scheduler with oracle knowledge of the application states. `Ideal' has no migrations cost and thus, exhibits the effect of solely contention for cores. Migration cost is 1 millisecond in the `Cost' and `Inertia' configurations. This cost exceeds reported costs on big.LITTLE machines~\cite{migration-biglittle} and recent work~\cite{migration-rewriting1, migration-rewriting2}. The `Inertia' configuration has a scheduler that uses inertia: applications that migrate are held to the assigned core for 5 schedulings. Additionally, we study the impact of higher migration costs to exhibit the stability of achieved speedups in `Inertia+5ms' and `Inertia+9ms', respectively having 5ms per migration and 9ms per migration.

\paragraph{Key Finding:} \textit{In each system configuration, Inertia achieves close to Ideal speed up. Additionally, the system is robust against high migration cost achieving within 1.5\% of the ideal speed up in extreme migration cost scenarios.} \textbf{This means, for single threaded applications, that specialization is a road ahead for performance improvement.}

\paragraph{Analysis:} First, realistic scheduling loses less than 1\% performance compared to oracle scheduling. The schedules where oracle knowledge outperforms realistic scheduling are at the state transitions. These account for 2\% of the total schedules. In all configuration and constraint pairs, applications execute on a core that matches their current stressed component approximately 70\% of the time. This provides explanation for how each constraint achieves relatively equal speed up. Furthermore, the Inertia configuration recovers more than half of the loss from introducing migration overheads. This is the goal of including inertia in the scheduler.

\section{Related Work}
\label{sec:related}
This section details the related work along two avenues. First, how this paper compares to prior phase analysis work. Second, we list some works that already find performance gain by adjusting systems for heterogeneous hardware. 

\subsection{Phase Analyses}
The classification and prediction of program phases has been a combed through topic applied to power management~\cite{analysis-static-range,analysis-superfine, analysis-newcounters} and security~\cite{analysis-sidechannel,analysis-cluster-security}. Machine learning has sparked new research especially for prediction~\cite{analysis-machinelearning2}. We refer the reader to recent surveys for a compendium of techniques~\cite{analysis-survey2020,prediction-survey2019}. This section presents the limitations of these works and how this work differs.

Isci et al. classifies two hardware performance counters through static windows identifying different boundedness phases: from CPU-bound to memory-bound. These phases lie on one axis and cannot represent as rich a space as is presented in our work. The phases inform prediction, which in turn informs DVFS. The system improves energy-delay product over the baseline by 18\% on average. Similarly, Bui and Kim examined super fine grain program phases with application to DVFS~\cite{analysis-superfine}. The characterization is at cycle-count granularity. Intervals can be 1 to 100K cycles which is 1ns to 100$\mu$s assuming a 1GHz frequency: much finer granularity than our work; however, it is not feasible to collect entire traces from large benchmarks like those in the SPEC CPU 2017 suite using their methodology. They ran MiBench benchmarks on a processor model on an FPGA. They find that short duration super fine granularity phases detect and adjust V/F levels for more power savings: between 1.5 to 2x savings. Overall, these works provide benefit via DVFS configuration; neither migrates programs to cores more suitable for the performance/watt expected in the detected phase. In contrast, a core idea of our work is migration based on phase behavior.

More complex classifiers and predictors have been proposed. Alcorta et al. use hardware performance counters to train a phase classifier and phase predictor both in the single core \cite{analysis-machinelearning1} and multicore cases \cite{analysis-machinelearning2, prediction-forecasting}. The single core work achieved higher accuracy than table based methods in phase classification. It also determined that for the combined classification and prediction task, two-level k-means clustering was the best classifier for the majority of predictors. In the multicore case, classification is done to separate data into multiple predictors - each specialized to a phase. The results shows that phase-aware LSTM had the best prediction with 23\% mean absolute percent error averaged across the benchmarks studied. Due to their complexity, these phase analysis tools increase analysis overhead therefore increasing the difficulty in initial implementation in a system setting. Our work demonstrates that simple methods are enough to substantially improve performance. 

\subsection{Support for Heterogeneous CMP}
Various areas have gained from exploiting heterogeneous CMP. Cao et al. analyzed VM services and found opportunity for power and performance gains through placing JIT, garbage collection, and the interpreter on a different smaller core \cite{support-yin-yang}. Alcorta et al. applied phase analysis and machine learning to choosing the L2 prefetcher in multicore chips \cite{support-prefetchers}. The dynamic selection of the prefetcher improved IPC by 5.8\% on average. Both demonstrate that using heterogeneous CMP offers opportunity for improvements when the systems built on top of the architecture acknowledge the heterogeneity.

\section{Conclusion}
\label{sec:conclusion}
SAHM introduces a principled, empirical approach to identifying opportunities for architectural specialization based on fine-grained performance counter monitoring. By mapping runtime behavior to a discrete set of states and designing specialized cores accordingly, SAHM pushes the frontier of single-thread performance without requiring speculation or instruction-level parallelism. Our exhaustive modeling shows that even modest specializations can yield substantial gains, offering a compelling path forward for future CPU designs.

\clearpage
\bibliographystyle{plain}
\bibliography{references}

\appendix

\section{Appendix}
\label{sec:appendix}
This figure shows the distribution of potential speed ups given a configuration. Each graph is associated with a branch configuration while each vertical gridline has the rest of the configuration listed on the x-axis. For example, the leftmost boxplot of the bottommost graph is the configuration with a baseline core and a branch specialty core that provides a speed up of 30\%.
\begin{figure*}
  \centering
  \includegraphics[width=\textwidth]{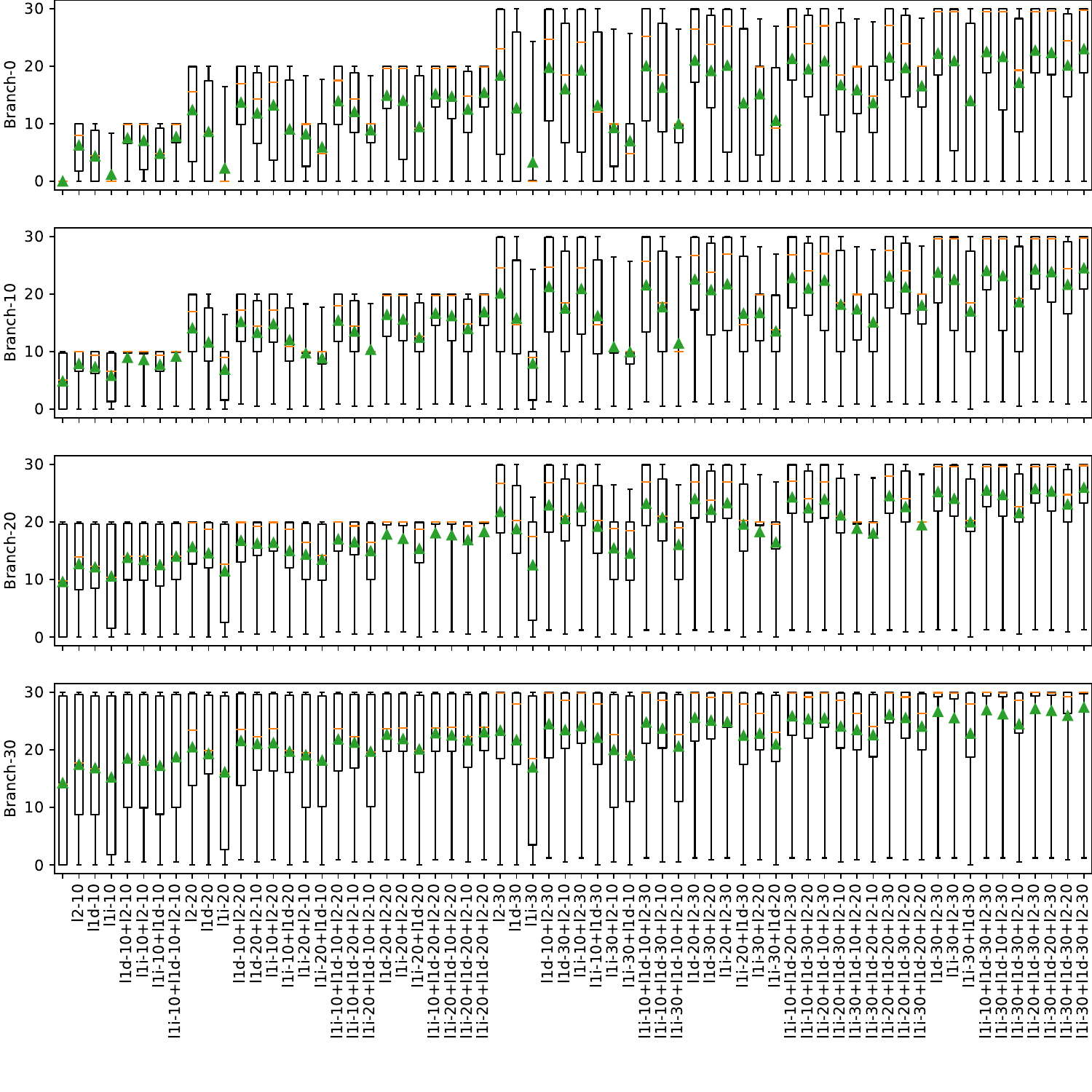}
  \caption{Distribution of application speed up by configuration. The whole design space is listed with the branch specialized core broken out into a chart each while the the rest of the cores are listed on the x-axis in ascending amount of speed up. Green triangle is mean while orange line is median. Box is 25\% and 75\% percentiles and the whiskers are the entire range.}
  \label{fig:complete_config_speedup}
\end{figure*}

\end{document}